

\input harvmac

\def\quarter{{\textstyle{1\over4}}} 
\def\ignore#1{}
\parindent 25pt
\parskip 20pt
\baselineskip 13pt
\hfill{ Brown HET-863; CERN-TH.6611/92;  HUTP-92/A035; LPTHE-Orsay: 92/29}
{\nopagenumbers
\vskip .3in
\centerline{\titlefont Symmetry Breaking In the Double-Well}
\vskip .1in
\centerline{\titlefont Hermitian Matrix Models}
\abstractfont
\vskip .3in

 \centerline{{\bf Richard. C. Brower}\footnote{$^\dagger$}{Physics Department,
Boston University, Boston, MA 02215; Theory Division, CERN, CH-1211, Geneva 23,
Switzerland; (brower@buphy.bu.edu).},{\bf Nivedita Deo}\footnote{$^\star$}{Mary
Ingraham Bunting Institute, Radcliffe/Harvard University, Cambridge, MA 02138;
Lyman Laboratory of Physics, Harvard University,
Cambridge, MA 02138; (after 9/1/92, Physics Department, Indian Inst.of Science,
Bangalore 560012, India).},{\bf Sanjay Jain}\footnote{$^\diamond$}{Lyman
Laboratory of Physics, Harvard University, Cambridge, MA 02138;  (after
9/1/92, Centre for Theoretical Studies, Indian Inst.of Science, Bangalore
 560012, India).},
 {and}
 {\bf Chung-I Tan}\footnote{$^\circ$}{Physics
Department, Brown University, Providence, RI 02912; Theory Division, CERN,
CH-1211, Geneva 23, Switzerland; (tan@brownvm.brown.edu).}}

\vskip .3in
\centerline{\bf Abstract}

We study symmetry breaking in $Z_2$ symmetric large $N$ matrix models.
In the planar approximation for both the symmetric double-well
$\phi^4$ model and the symmetric Penner model, we find there is an
infinite family of broken symmetry solutions characterized by
different sets of recursion coefficients $R_n$ and $S_n$ that all lead
to identical free energies and eigenvalue densities.  These solutions
can be parameterized by an arbitrary angle $\theta(x)$, for each value
of $x = n/N < 1$. In the double scaling limit, this class reduces to a
smaller family of solutions with distinct free energies already at the
torus level. For the double-well $\phi^4$ theory the double scaling
string equations are parameterized by a conserved angular momentum
parameter in the range $0 \le l < \infty$ and a single arbitrary
$U(1)$ phase angle.

\vfill
\noindent  August 1992
\hfill
\eject

}

\pageno=1
\baselineskip=13pt plus 2pt minus 1pt

\newsec{Introduction}
We would like to highlight some unusual  aspects of symmetric
double-well matrix models
\ref\multicuti{Y. Shimamune, Phys. Lett. {\bf  B108} (1982) 407;
G.M. Cicuta, L. Molinari, and E. Montaldi, Mod. Phys. Lett.
{\bf  A1} (1986) 125, J. Phys {\bf  A23} (1990) L421;
J. Jurkiewicz, Phys. Lett. {\bf 245} (1990) 178;
G. Bhanot, G. Mandal, and O. Narayan, Phys. Lett. {\bf B251} (1990) 388.}%
\nref\DDJT{K. Demeterfi, N. Deo, S. Jain, and C-I Tan, Phys. Rev
{\bf  D42} (1990) 4105.}%
\nref\multicutii{L. Molinari, J. Phys. {\bf A21} (1988) 1;
O. Lechtenfeld, R. Ray, and A. Ray, Int. J. Mod. Phys. {\bf A6} (1991) 4491.}%
\nref\DSS{M. Douglas, N. Seiberg,  and S. Shenker, Phys. Lett. {\bf B244}
(1990) 381.}%
\nref\CM{C. Crnkovic and G. Moore, Phys. Lett. {\bf B257} (1991) 322.}%
\nref\MS{P. Mathieu and D. Senechal, Mod. Phys. Lett. {\bf A6} (1991) 819.}%
\nref\Nappi{C. Nappi, Mod. Phys. Lett. {\bf A5} (1990) 2773.}%
\nref\Pi{P.M.S. Petropoulos, Phys. Lett. {\bf B261} (1991) 402.}%
\nref\CDM{C. Crnkovic, M. Douglas, G. Moore, Yale and Rutgers preprint
YCTP- P25-91, RU-91-36.}-\ref\HMPN{T. Hollowood, L. Miramontes, A.
Pasquinucci, and C. Nappi, Nucl. Phys. {\bf B373} (1992) 247.}
concerning spontaneous symmetry breaking and the multiplicity of
solutions at the same critical point.  We consider two specific
models: (i) a $\phi^4$ theory with a symmetric double well and (ii)
the symmetric Gaussian Penner model
\ref\Ti{C-I Tan, Mod. Phys. Lett. {\bf A6} (1991) 1373.}%
\nref\Tii{C-I Tan, Phys. Rev. {\bf D45} (1992) 2862.}%
-\ref\CDL{S. Chaudhuri, H. Dykstra, and J. Lykken, Fermilab preprint
FERMI-CONF-91/190-T, to appear in the proceedings of the XXth International
Conference on Differential Geometric Methods in Theoretical Physics, New York,
June 1991.}. Both models have a $Z_2$ reflection symmetry and a standard
two-band solution that respects it. However we obtain new classes of solutions
that break the $Z_2$ symmetry by relaxing the initial boundary conditions on
the
first two recursion coefficients for the othogonal polynomials.

The single-band broken symmetry solution to the double-well $\phi^4$
theory is c=0 pure gravity, but at lower free energy there is an
infinite class of two-band solutions where
 the tree level eigenvalue density is symmetric
in  the two wells. These solutions have the property that
their free energy
and eigenvalue density, in the planar limit, are invariant with respect to
an infinite set of continuous parameters in the recursions
coefficients.  An analgous two-band class of broken symmetry solutions
is found in the Penner model as well.  In the double scaling limit for
the $\phi^4$ theory the degeneracy is lifted, except for a single U(1)
rotation, and a one parameter family of solutions survives satisfying
the Painleve-II equation with an extra conserved ``angular momentum''
parameter. At this stage the physical consequences of this degeneracy
of solutions are not clear. Since the tree level eigenvalue density is
the same for all these solutions it is possible that some of them
could tunnel into each other with an instanton action that is lower
order in $N$. If so, this would be  of particular interest  in the context
of the Gaussian Penner model, where one of the broken symmetry solutions
discussed is related to the $c=1$ string at twice the self-dual radius.

The organization of this paper is as follows: In section 2, we give a
brief overview of symmetry breaking and the formalism for the
orthogonal polynomial method. In section 3, we consider the
consequences of relaxing the boundary condition on the recursion
coefficients for the symmetric $\phi^4$ potential and the Gaussian
Penner model, and classify all two-cut solutions in the planar ($N
\rightarrow \infty$) limit. In particular we exhibit the class of
solutions that have the same tree level eigenvalue density and free
energy. We discuss some correlators that distinguish between the
various solutions and a numerical approach to investigating finite N
solutions. Section 4 discusses the double-scaling limit of the free
energy for both models emphasizing that one gets an expanded class of
double-scaling solutions.  The discussion and conclusions are given in
section 5.

\vfill\eject
\newsec{Overview of Problem}

It has been suggested that the singular behavior of the tree level
eigenvalue density near the edge of the cuts determines the critical
behavior of the matrix model in the following sense
\ref\evlore{V.A. Kazakov, Mod. Phys. Lett. {\bf A4} (1989) 2125;
D. Gross and A.A. Migdal, Nucl. Phys. {\bf B340} (1990) 333; H.
Neuberger, Nucl. Phys. {\bf B352} (1991) 689; S. Dalley, C. Johnson,
and T. Morris, Phys. Lett. {\bf B262} (1991) 18, Mod. Phys. Lett.
{\bf A6} (1991) 439; M. Bowick and E. Brezin, Phys. Lett. {\bf B268}
(1991) 21.}\CM\CDM: One identifies the polynomials in the matrix
variable $\phi$ or scaling operators $O_n$ ($O_n$'s are traces of the
polynomials) which produce a particular kind of singularity (labeled
by $n$) near the edge of the cut or cuts. Knowing the $O_n$'s, one
then considers $Z(t)=\int d\phi \exp[-\sum t_n O_n(\phi)]$ and shows
 that $Z(t)$ satisfies a certain hierarchy of equations.
This hierarchy depends, then, only on the kind of ensemble of matrices
considered (hermitian, antihermitian, unitary, etc.), the class of
singularities of the eigenvalue density allowed (e.g., single-cut
density with multiple zeroes coalescing at the edge, two cuts
colliding and sandwiching zeroes in between, etc.), and any symmetry
of the potential that restricts the class of eigenvalue densities
considered (e.g., restriction to even perturbations in $Z_2$ symmetric
2-cut models)
\HMPN.

{\noindent \it Orthogonal Polynomials}

It is generally believed  for potentials $V(\phi)= t_n O_n$ that
are bounded from below that the orthogonal polynomial method
\ref\BIZ{D. Bessis, Comm. Math. Phys. {\bf 69} (1979) 147;
D. Bessis, C. Itzykson, and J. B. Zuber, Adv. Appl. Math. {\bf 1}
(1980) 109.}  uniquely fixes the
solution for the free energy, correlators, etc. Here we
will show that orthogonal polynomial method actually allows one to
construct a whole class of closely related solutions.  To clarify this
further,  we remind the reader of the precise condition for a unique
solution for the free energy in terms of orthogonal polynomials.
Consider the partition function,
${Z_N=\int d\phi~~ e^{-N tr V(\phi)},}$
where $\phi$ is an $N \times N$ hermitian matrix and $V(\phi)$ is a real
potential. The integral is  expressed
in terms of  a set of  orthogonal polynomials $P_n(x)$,
$\int dx~ P_n(x) ~ P_m(x) ~e^{-N V(x)} = h_n \delta_{nm},$
normalized by the convention that the leading term for $P_n(x)$ is $x^n$,
${P_n(x)=x^n+c_1 x^{(n-1)}+......}$.
(Note that this convention sets $P_0(x) = 1$ or $h_0 =\int dx~ exp{[-N
V(x)]}.$) Since these orthogonal polynomials can be iteratively determined by
Gram-Schmidt orthogonalization, the exact free energy, $F_N = \log Z_N =
\sum_{n = 0}^{N-1} \log h_n,$ as well as all thermodynamic averages are {\it
uniquely} determined.

Instead of actually finding these orthogonal polynomials, one in practice  uses
recursion relations for the coefficients $R_n$ and $S_n$ in the expression, $x
P_n(x)=P_{n+1}(x)+S_n P_n(x)+R_n P_{n-1}(x)$. Once the $R_n$'s are known, the
 free
energy can be found by using the fact that $R_n=h_n/h_{n-1}$. It is convenient
to introduce a self-dual orthonormal basis $|n\rangle$, where $\langle
 x|n\rangle
= P_n(x)/\sqrt{h_n}$, and $|x\rangle$ are eigenvectors of the operator $\hat
\phi$ with eigenvalues $x$,  satisfying the normalization $\langle x'|x\rangle
=
exp(N V(x) )$ $\delta(x'-x)$.  Matrix elements of the operator ${\hat \phi}$
in this new orthonormal  basis are  directly related to the recursion
coefficients by $\langle m|{\hat
\phi}|n\rangle=\sqrt{R_m}\delta_{m,n+1}+S_n\delta_{m,n}
+\sqrt{R_n}\delta_{m,n-1}.$  In terms of ${\hat \phi}$, the recursion relations
for $R_n$ and $S_n$ can be expressed in operator notation,
\eqn\rseq{\eqalign{n/N&=\sqrt{R_n} \langle n-1|V'({\hat \phi})|n\rangle\cr
0&=\langle n|V'({\hat \phi})|n\rangle.\cr}}
Once initial conditions are specified, a {\it unique} solution for
$R_n$ and hence the normalization for each orthogonal polynomial,
$h_n$, can be found by iteration. For example, in the case of the $\phi^4$
 model,
Eq. \rseq\ is a pair of coupled two-term recursion equations for $R_n$ and
$S_n$
which require four inital conditions:  the numerical values
\eqn\rseq{\eqalign{S_0 &= \langle
0|{\hat \phi}|0\rangle={h_0}^{-1}\int dx x e^{-NV(x)}\cr
R_1 &= \langle 0|{\hat \phi^2}|0\rangle -
\langle 0|{\hat \phi}|0\rangle^2= {h_0}^{-1}\int dx x^2 e^{-NV(x)}-S_0^2 ,\cr}}
and the trivial values $R_0=0$ and $S_{-1} = 0$, which are independent of
the potential. Given these values, all other coefficients are given
iteratively as rational function of $S_0$ and $R_1$.  (A more elegant
formulation would introduce  a single sequence of coeffients $C_k \propto
\langle 0|{\hat \phi^k}|0\rangle + ....$  with a single four-term recursion
relation $C_{k} = F_k(C_{k-1},C_{k-2},C_{k-3},C_{k-4})$, where $C_{k} =
R_{k/2}$ for k even and $C_{k} =S_{(k-1)/2}$ for k odd.)

{\noindent \it Symmetry Breaking}

This formalism  should make it clear that there is {\it no} ambiguity in
 defining the
matrix models for all {\it finite} $N$, assuming of course that the potential
is
{\it bounded from below} and that the integrals defining the inner product are
finite. However typical of all statistical mechanical problems this does not
imply that we know the correct way to take the thermodynamic (or in this
instance large $N$) limit. To understand this potential source of ambiguity in
using the
 recursion
relations at large $N$, consider the double-well potential,
${V(\phi)=\sigma \phi + \half \mu ~\phi^2+\quarter g ~\phi^4}$, with
$\mu<0$, $g>0$ and a small symmetry breaking term $\sigma \phi$.
To investigate symmetry breaking it is useful to study the effect of
interchanging the limits $N \rightarrow \infty$ and $\sigma
\rightarrow 0 \pm $ on the values of $S_0 = \langle 0|\hat \phi|0 \rangle$.
By $Z_2$ symmetry, if we take $\sigma \rightarrow 0\pm$ followed by $N
\rightarrow \infty$, we must get $S_0 = 0$, whereas if we take $N
\rightarrow \infty$ followed by $\sigma \rightarrow 0\pm$
we  have
\eqn\szeroone{S_0 = \pm \sqrt{-\mu/g},}
as can be demonstrated by using steepest descent at the stationary
minima of the potential $V(x)$.  Similarly, to complete  the necessary boundary
conditions, one can show  that $R_1 = \langle 0|\hat \phi^2 |0 \rangle - S_0^2$
takes on   $R_1 = {-\mu/g}$  and $R_1=0$ for these two limits respectively.
Now we can in principle use the recursion relations to obtain both symmetric
and
broken symmetry solutions. In general terms, this is just the familiar feature
of spontaneous symmtry breaking. For example the Ising model with no external
field on a finite lattice must  have $\langle s_i\rangle =
0$ by $Z_2$ symmetry, but in the large volume limit (at temperatures
below the Curie point) the relevant ({\it i.e.} stable) solution is a broken
vacua with non-zero values for $\langle s_i\rangle = \pm m$ obtained by
applying
an infinitesimal magnetic field.

As we will see shortly, a more general possibility  at {\it infinite $N$} is
to characterize the ``vacuum'' state for the double-well $\phi^4$ model  by a
mixing angle $\theta_0$, %
\eqn\vac{| 0, \theta_0 \rangle = cos(\theta_0/2) \; | 0, + \rangle +
 sin(\theta_0/2)\; | 0, - \rangle}
where $ | 0, \pm \rangle$ are
orthonormal and $\hat \phi | 0, \pm \rangle \simeq \pm \sqrt{-\mu/g} |
0, \pm \rangle$. As a consequence for the mixed state the  boundary conditions
 becomes
\eqn\szero{ S_0 = \sqrt{-\mu/g}\;cos(\theta_0),}
instead of \szeroone, with the  constraint,
\eqn\constraint{ R_1 = -\mu/g - S_0^2 \ge 0.}

The goal of this paper is to understand how the solutions of the
double-well matrix models depend on the initial boundary conditions,
{\it e.g.}, for the $\phi^4$ model, on the first two moments, $S_0$
and $R_1$.  In the large $N$ (or planar limit) we find a large class
of solutions consistent with the broken symmetry boundary condition in
addition to the (meta stable) pure gravity solution at higher free
energy.  Indeed we will show that the above qualitative discussion is
a rigorous consequence of the planar solutions in the two-band ansatz.
More generally however, we feel that the lack of a full understanding
of the effects of this boundary condition represents an important gap
in our ability to fully determine and solve the string equations
resulting from matrix models.
\vfill
\break\eject

\newsec{Multiple Solutions In Matrix Models : Tree Level Analysis}

In this section we establish the existence of multiple solutions in
two models: (i) the double well $\phi^4$ model and (ii) Gaussian Penner model.

\subsec{The Double Well $\phi^4$ potential}
For the double well potential ( ${V(\phi)=\sigma \phi + \half \mu
{}~\phi^2+\quarter g ~\phi^4}$), Eq. \rseq\ reduces to the recursion relations
\eqn\rseqmug{\eqalign{{n\over N}&=R_n [\mu+g
(R_{n+1}+R_n+R_{n-1}+S_n^2+S_{n-1}^2+ S_{n-1} S_n)]\cr 0&=\sigma + \mu S_n+g
[R_{n+1} (S_{n+1}+2 S_n)+R_n (2 S_n+S_{n-1})+S_n^3].\cr}}
We shall  first illustrate our procedure by considering a
symmetry-breaking solution under a period-one ansatz for both the R's and
S's: $R_n\rightarrow R({n\over N})$, $S_n\rightarrow S({n\over N})\neq 0$.
 Ignoring the $1/N$ corrections, Eq. \rseqmug\ leads to two relations, which
allow us to solve for $R(x)$ and $S(x)$:
\eqn\gravity{\eqalign{R(x) &=(1/15g)[-\mu-\sqrt{\mu^2-15gx}]\cr
S(x) &=\pm\sqrt{-\mu/g-6R(x)}.\cr}}
With $\mu<0$ and $g>0$, one has $R(0)=0$, $S(0)=\pm\sqrt{-\mu/g}$, consistent
with our discussion on symmetry breaking in Sec. 2, and $R(x)$ monotonically
increasing for $0<x<\mu^2/15g$.

The generating function $F(z)\equiv {1\over N}\langle Tr{1\over
z-\phi}\rangle$ for a period-one ansatz at the tree level can be
expressed in terms of $R(x)$ and $S(x)$ as follows:
\eqn\genfunoneband{F(z)=\int_0^1dx\int_0^{2\pi}{d\theta\over
2\pi}{1\over z-\phi(x,\theta)}=\int_0^1dx{1\over \sqrt{[z-S(x)]^2-4R(x)}},}
where $\phi(x,\theta)=S(x)+\sqrt{R(x)}(e^{i\theta}+e^{-i\theta})$. One can
 verify
by explicit calculation using \gravity\ that the eigenvalues lie in an interval
$[z_-,z_+]$, where $z_{\pm}\equiv S(1)\pm 2\sqrt{R(1)}$, with
a single-band  eigenvalue density, $\rho(z)$, in
agreement with the result of Shimamune (Ref. \multicuti) obtained by using the
Schwinger-Dyson equation.  On the line $\mu=-\sqrt{15g}$, $R(x)$ develops a
 square-root type singularity at $x=1$, leading to a Painleve-I equation in the
double scaling limit, appropriate for the $c=0$ 2D gravity solution.

Since all the eigenvalues are concentrated in a single well, this
symmetry-breaking solution does not correspond to a configuration with the
 lowest
free energy, but is a subdominant solution. In the double scaling limit it is
unstable against the tunnelling of a single eigenvalue into the other well,
exactly like the subdominant solution for the pure gravity in the $\phi^6$
model. We now turn to solutions which have eigenvalues in both
wells, which include a class of solutions for which $\rho(z)$ is exactly
$Z_2$-symmetric (when $\sigma=0$) at the tree level, but which in general break
this symmetry at higher order in ${1/N}.$

Let us  consider  a
period-two ansatz for both the R's and S's
\eqn\rsabcd{\eqalign{R_n&=A({n\over N})
\; , \;\; S_n=C({n\over N}) {\rm~~for~n=even}, \cr
R_n&=B({n\over N}) \; , \;\;  S_n=D({n\over N}){\rm~~for~n=odd}.\cr}}
Taking $A$, $B$, $C$ and $D$ to be continuous, and ignoring the $1/N$
corrections in the recursion equations \rseqmug, we obtain four tree level
recursion relations,
\eqna\abcd$$\eqalignno{2x&=\mu_{eff} (A+B)+ g (A^2 + 4AB +B^2), &\abcd a\cr
0&=(A-B)(A + B+ {\mu_{eff}\over g}), &\abcd b\cr 0&=2\sigma +\mu (C+D) +g [3
(A+B)(C+D) + C^3 +D^3 ],&\abcd c\cr 0&=(C-D)(A + B+ {\mu_{eff}\over g}),&\abcd
d\cr}$$
where $\mu_{eff}\equiv \mu +g(C^2+CD+D^2)$.

Since we have already considered the single-band (pure gravity
case), we can  assume that either $A \ne B$ or $C \ne D$. To carry out an
exhaustive analysis of the full set of solutions to these equations, it is
useful to note that only three out of the four equations are independent. The
three independent equations take the simple form,
\eqn\simple{{A+B-CD= -{\mu\over g}-(C+D)^2 ~, ~ A  B ={ x\over g}  ~,~~{\rm
and}~ V'[-(C+D )] = 0.}}
Note that there is no condition on $C-D$ which can be independently chosen
for every value of $x$. The first two equations allow one to express the
 explicit
solution for $A$ and $B$ in the familiar form\DDJT, %
\eqn\symmAB{ A ={1\over 2g}( - \mu_{eff} \pm \sqrt{\mu_{eff}^2 -4gx}~)
 \hskip30pt B = {1\over 2g}(- \mu_{eff} \mp \sqrt{\mu_{eff}^2 -4gx}~).}
In general for the double-well potential (for $\mu < 0$ and small
enough $\sigma$), there are three $x$-independent real solutions to $V'[-(C+D
)]
= 0$.  With $C+D$ fixed, it
follows that the first combination $A+B-CD$ in \simple\ is also fixed.
Consequently, a ``circular'' constraint on  $A-B$ and $C-D$ can be found
\eqn\circleconstraint{(A-B)^2 + {4 x \over g} = {1 \over 16}[ (C-D)^2 +{4 \mu
\over g} + 3(C+D)^2]^2.}

This constraint can be represented by contours for each fixed value of
$x$, $0\leq x\leq 1$, in a two-dimensional plane with $A-B$ as the
vertical axis and $C-D$ the horizonal axis. In \fig\merdeglace{
Constraint on $A-B$ versus $C-D$ for $x=0,\>0.5,\>.75,\>.875,\>.95$,
with $\mu=-2$ and $g=1$.}, we exhibit them for the class of solutions
where $\sigma=0$ and $C+D=0$.  The external contour, corresponding to
$x=0$, is precisely the constraint suggested in the qualitative
discusion of symmtry breaking of Sec. 2. To see this one must consider
carefully the proper definitions at the boundary, namely $A - B = R_0
- R_1 = - R_1$ and $C = - D = S_0$ at $x=0$, which yields the
contraint Eq.
\constraint\ in the form,
$$R_1^2 = ( S_0^2 + \mu/g)^2.$$
For $x=x_{cr}\equiv \mu^2/4g$, the contour shrinks to a point, $A-B=C-D=0$,
about
which a double scaling limit can be taken \ref\critical{For
$x>x_{cr}$, only the symmetric one-band solution, $A=B$ and
$C=D=0$, survives. Criticality therefore occurs at
$x_{cr}=1$, {\it i.e.},  $\mu=-2\sqrt g$.}.

The different solutions to \simple\ can be parametrized by curves in
this plane traversing from the $x=0$ to the $x=1$ contours (see for example
fig.
4(b)).  For instance, a solution can be specified, at each value of $x$, by the
polar coordinate, $\theta(x)$, for the intersection of the curve with
the contour
\circleconstraint.
  Conversely, once $\theta(x)$ is chosen for every $x$, a unique solution to
Eq.
\simple, ($A(x), B(x), C(x), D(x)$), is obtained \ref\merparam{An
alternative parametrization is, $\quarter( C-D)^2 = (-\mu /g) \;
cos^2\theta(x)$ and $(A-B)^2 = (-\mu/g)^2 \; sin^4\theta(x) - 4 x/g$,
from which both the $x=0$ and $x=1$ contours can be readily obtained.
In this parametrization, $\theta(x)$ is restricted to
$-\theta_{max}\leq\theta(x) \leq \theta_max(0)$ where
$sin^4\theta_{max} = 4 g x/\mu^2$ and $\theta_{max}$ goes to $\pi/2$
at $x_{cr}$. Another convenient parametrization which turns all
contours into concentric circles is, $Y=r(x) sin \theta(x)\equiv A-B$,
$X=r(x) cos\theta(x)\equiv {\rm sign}(cos
\theta)[(-\mu/2g)(C-D)^2-(C-D)^4/16]^{-1/2}$, where
$r^2=(\mu/g)^2(x_{cr}-x)$. In this parametrization, $\theta(x)$ is
unrestricted.}. In analogy with \vac, each state, $|n\rangle$, $0\leq
n/N\leq 1$, in the large $N$ limit could be thought of as a linear
conbination of ``left-" and ``right" states, specified by an arbitrary
density function, $ \theta(x)$.  The choice of $\theta(x)$ represents
the ambiguity of solution at the tree level. (In fact the ``orbit''
need not even be continuous.  However, if the orbit is discontinuous,
derivatives of $A, B, C, D$ are large and cannot be ignored, as is
assumed in the tree level analysis.)

An interesting feature of these broken symmetry solutions is that,
within the class specified by one of the three values of $C+D$, they
all have the {\it same} tree level eigenvalue density and free energy.
To see this consider the generating function $F(z)\equiv {1\over N}
\langle Tr {1\over (z-{\phi})}\rangle$ for a {\it general} period-two ansatz at
 tree
level\ref\gentwoband{This representation for a period-two structure
was first derived in Ref. \DDJT\ under the symmetric ansatz where
$C=D=0$. With $C\neq 0\neq D$, in analogy to Eq. \genfunoneband, one
can introduce a $2\times 2$ matrix $\Phi_{\alpha,\beta}(x,\theta)$,
where $\alpha, \beta$ take on values 0 and 1, (for even and odd
respectively), i.e., $\Phi_{0,0}=C(x)$, $\Phi_{1,1}=D(x)$,
$\Phi_{0,1}=\sqrt A e^{i\theta}+\sqrt B e^{-i\theta}$, and
$\Phi_{1,0}=\sqrt A e^{-i\theta}+\sqrt B e^{i\theta}$. It follows that
$F(z)=1/2\int_{0}^1 dx\int_0^{2\pi}{d\theta\over
2\pi}\sum_{\alpha}[z-\Phi(x,\theta)]^{-1}_{\alpha,\alpha}$. The
$\theta$-integration can easily be carried out.} %
\eqn\gen{F(z)=1/2\int_{0}^1
dx{(2z-(C+D))\over\sqrt{[z^2-z(C+D)-(A+B-CD)]^2-4AB}}.}
 Notice that Eq. \gen\
involves precisely the three combinations, Eq. \simple, which are fixed by the
four recursion relations. Therefore, once the value of $C+D$ is chosen,
one gets the same generating
function $F(z)$. It follows that the tree level  eigenvalue density
$\rho(x)$  and  free energy are the {\bf same} for all solutions within a
class, {\it i.e.}, are independent of the choice of $\theta(x)$.

For the most part we now shall restrict further discusions to the case
$\sigma= 0$ and $C + D = 0$. All the solutions in this class give rise
to the same eigenvalue density $\rho(z)$. This class of solutions is
continuously deformable into the limiting case $A(x) \neq B(x)$,
$C(x)=D(x)=0$, which is just the standard {\it symmetric} solution for
the two-band solution, with $\mu_{eff}=\mu$.  At the other extreme
there is a {\it maximally asymmetric} two-band solution satisfying the
condition $A(x)=B(x)$, $C(x)=- D(x)\neq 0$, with
\eqn\asym{A(x)=B(x)=R(x)=\sqrt{x/g}, \hskip30pt
C(x)=-D(x)=\pm\left[{|\mu|/g}-\sqrt{4x/g}\right]^{1/2}.} These values
of $C$ and $D$ form the turning points at which our numerical
solutions change the sign of $A-B$. (See fig. 1.) The symmetric
solution corresponds to the choice $\theta(x)=\pm \pi/2$ and the
maximally asymmetric solution to the choice $\theta(x)={0,\>\pi}$.
Our expanded class of solutions includes ones where the branch of the
square root singularity ($\pm$) in Eq. \symmAB\ is exchanged between
$A$ and $B$ as the trajectory rotates in the $A-B$ versus $C-D$ plane,
as we note in the discussion of our numerical
results for finite $N$ (see Sec 3.4 and fig. 4 (c)). This corresponds to
$\theta(x)$ winding around the circle a number of times as $x$ goes
from $0$ to $1$. This is a precursor of the angular momentum variable
of the double scaling solutions. The rigid implementation of $Z_2$
symmetry and the boundary conditions on the recursion coefficients
would have yielded only the $\theta(x)=\pm \pi/2$ solution. All other
solutions correspond to a breaking of the $Z_2$ symmetry.

\subsec{Gaussian Penner Model}
The second example, we would like to consider, is the Gaussian Penner model.
The
potential for a general Penner model is ${V(\phi)=V_0(\phi)-t ~~log ~\phi}
$, where $V_0$ is a polynomial\Tii. If $V_0(\phi)=\phi$ the
model is the linear Penner model
\ref\Penner{J. Harer and D. Zagier, Invent. Math. {\bf 85} (1986) 457;
R.C. Penner, Bull. Am. Math. Soc. {\bf  15} (1986) 73,
J. Diff. Geom, {\bf 27} (1988) 35.}\ref\DV{J. Distler and C. Vafa, Mod. Phys.
Lett. {\bf A6} (1991) 259.}, if $V_0(\phi)=\mu ~\phi^2/2$ the model is the
Gaussian Penner model \Tii\CDL, where we interpret the $\log\phi$ term as
${1\over 2}\log\phi^2$. In \fig\pen{The Gaussian Penner potential for $t< 0$,
$t=0$ and $t> 0$.} we display the Gaussian Penner potential for different
values
of t. Consider first the situation where $t>0$. (The region $t<0$ is reached by
analytic continuation.\Ti\Tii) As the potential is a double well, the
period-two ansatz may be applied here also.

The recursion relations (Eq. \rseq) for a general Penner model reduce
to
\eqna\rseqpen$$\eqalignno{{n\over
N}&=\sqrt{R_n}\langle n-1|V_0^\prime({\hat \phi})|n\rangle -t\sqrt{R_n}\langle
n-1|{\hat \phi}^{-1}|n\rangle&\rseqpen a\cr 0&=\langle n|V_0^\prime({\hat
\phi})|n\rangle-t\langle n|{\hat \phi}^{-1}|n\rangle.&\rseqpen b\cr}$$
Let us denote $W_n=\sqrt{R_n}\langle n-1|V_0^\prime({\hat \phi})|n\rangle$ and
$Y_n=\langle n|V_0^\prime({\hat \phi})|n\rangle$ for later notational
convenience. For the Gaussian Penner model, $W_n=\mu R_n$ and $Y_n=\mu S_n$.

Eqs. \rseqpen{a,b}\ are unusual since they involve matrix elements
of $\hat\phi^{-1}$. For $t>0$, they can be solved in
the spherical limit by a procedure similar to that used for deriving the
generating function $F(z)$,  Eq.
\gen, under a period-two ansatz \ref\phiinverse{ The
matrix elements of $\hat \phi^{-1}$ can be expressed in terms of $A(x)$,
$B(x)$,
$C(x)$ and $D(x)$ by working with  the $2\times 2$ matrix $\Phi(x,\theta)$
mentioned in Ref. \gentwoband. Introduce two convenient combinations:
$\eta\equiv A(x)+B(x)-C(x)D(x)$ and $\xi\equiv A(x)B(x)$. One finds that
$\sqrt{R_n}\langle n-1|{\hat \phi}^{-1}|n\rangle\rightarrow
1/2+(A-\eta/2)/\sqrt{\eta^2-\xi}$, $\langle n|{\hat
\phi}^{-1}|n\rangle\rightarrow -D(x)/\sqrt{\eta^2-\xi}$ for $n$ even, and
$\sqrt{R_n}\langle n-1|{\hat \phi}^{-1}|n\rangle\rightarrow
1/2+(B-\eta/2)/\sqrt{\eta^2-\xi}$, $\langle n|{\hat
\phi}^{-1}|n\rangle\rightarrow -C(x)/\sqrt{\eta^2-\xi}$ for $n$ odd.}. By
considering $n$ even and odd, Eqs. \rseqpen{a,b}\ should normally lead to four
conditions.  Just like the $\phi^4$ model, only  three are
independent, and they can be cast in the following form:
\eqna\abcdpen$$\eqalignno{C+D&=0,&\abcdpen a\cr
           A+B-CD&={2x+t\over \mu},&\abcdpen b\cr
           AB&={x(x+t)\over \mu^2}.&\abcdpen c\cr}$$
For the symmetric solution where $C = D = 0$, one finds
\eqn\absympen{A(x)={x\over \mu}, \hskip 40pt B(x)={x+t\over
\mu}.} For the maximally asymmetric solution, on the other hand, one has
\eqn\acasym{A(x)=B(x)={1\over\mu}\sqrt{x(x+t)},\hskip20pt
C(x)^2={1\over \mu}\left[(2x+t)-2\sqrt{x(x+t)}\right]. }

Observe that Eqs. \abcdpen{a-c}\ are precisely the necessary combinations which
are needed in Eq. \gen\ for determining the generating function of our
symmetric Gaussian Penner model in the spherical limit leading to  symmetric
two-band structure. Therefore, for this class of solutions and in particular
for
the symmetric and maximally asymmetric solutions,  the eigenvalue density and
free energy are again identical at tree level  \ref\puzzle{This also resolves
th
contradiction between Refs. \Tii\ and \CDL. These papers contain two
different solutions to the Gaussian Penner model at the same critical
point. However, from the viewpoint of the present work, this should
not be regarded as a contradiction, but rather a special case of a
general phenomenon in multi-cut matrix models -- the existence of
multiple solutions. The solutions obtained in Refs. \Tii\ and \CDL\
are the symmetric and maximally asymmetric solutions respectively.}.

\subsec{Correlation Functions that distinguish between symmetric and
asymmetric
ansatz solutions}

Given that the tree level generating function and hence eigenvalue
density and free energy are the same, one might ask if there are other
correlation functions that distinguish between the various solutions.
It is normally assumed that after taking the period-two ansatz the
large N limit is smooth for all correlators. However, consider the
correlator $\langle Tr\phi Tr\phi\rangle_c$. In terms of recursion
coefficients,
\eqn\phiphi{\langle Tr\phi Tr\phi\rangle_c=R_N.}
In the symmetric solution, since the $R_N$ alternate between $A_N$ and
$B_N$ as $N$ goes from odd to even, this correlator at large $N$
depends on whether $\infty$ is approached through odd or even $ N$. In
one case it is $A(1)$, in the other $B(1)$. For the $\phi^4$ model,
they differ even at tree level by $\sqrt{\mu^2-4g}/g$ which is
singular at criticality, (see Eq. \symmAB). On the other hand in the
maximally asymmetric solution $S_N$ is period two but $R_N$ is of
period one, (see below Eq. \asym) hence this correlator has no
discontinuity between odd and even. This is an example of a correlator
that distinguishes between the two solutions. A similar difference between odd
and even $N$ is known to exist in the context of unitary matrix models
\ref\Wadia{S.R. Wadia, Phys.  Lett. {\bf B93} (1980) 403.}.

Another example is
$\langle Tr\phi Tr\phi
Tr\phi\rangle_c=R_N\left(S_{N-1}-S_N\right).$
In the symmetric solution $S_N=0$ and this vanishes.   But in the maximally
asymmetric solution since $S_n$ is period two, (and $C=-D$), $S_{N-1}-S_N$
changes sign as one goes from odd to even $N$. In particular, $\langle Tr\phi
Tr\phi Tr\phi\rangle_c=(|\mu|/g^2-2({1/g})^{3/2})^{1/2}$ for $N$ odd and its
negative for $N$ even.

 Another  characterization of the difference between the solutions is the
following: If one truncates the infinite dimensional matrix  $\langle
n|{\hat \phi}|m\rangle$ to an $N\times N$ matrix corresponding to the subspace
of the first $N$ orthogonal polynomials, the eigenvalues of this $N \times N$
matrix are a good approximation to the saddle point configuration of the
eigenvalues at large $N$. Since its matrix elements are given in terms of
$R's$
and $S's$, we can determine the eigenvalues numerically from a given solution
of the recursion coefficients.
In  \fig\evood{Eigenvalue distributions for even
and odd $N$: (a) $N=24$, symmetric solution; (b) $N=24$, asymmetric solution;
(c) $N=25$, symmetric solution; (d) $N=25$, asymmetric solution.}, we show the
locations of the eigenvalues so obtained for the tree level symmetric and
maximally asymmetric solutions for $N=24$, $25$. For $N=24$, half the
eigenvalues are in one well and half in the other, for both solutions. However,
for $N=25$, (odd $N$), there is a striking difference between the two
solutions.
For the asymmetric solution, there is one extra eigenvalue located in one of
the
two wells, (the well selected depends upon the sign of $S_0$), whereas for the
symmetric solution, this extra eigenvalue in the the center (on top of the
barrier), thus preserving the symmetry between both the wells.

\subsec{Numerical Approach to Finite N Solutions}

Another approach to understanding the role of these multiple solutions
is to pursue a numerical study of finite $N$ solutions and attempt to
take $N$ large enough to see a cross over to the large $N$ (or double
scaling) regimes.  Although we will postpone a detailed analysis of
our results, there are several general features which can help to
understand the present discussion. The recursion relations for the
double well potential follow from the variation of an effective
action, \ref\Veff{ The recursion relations , Eq. \rseq, are the
Eulerean stationarity conditions for this effective action. This
approach to the finite N recursion relations was brought to our
attention by A.  Jevicki ( private communication).} %
\eqn\effect{\eqalign{V_{eff}(R_n,S_n)= \sum_{n = 0}^\infty  \{  - {n\over N }
lo
g(R_n)
&+ \mu R_n + { g\over 2} (R^2_n+ 2 R_n R_{n+1} )\cr
&+ \sigma S_n + {\mu \over 2} S_n^2 + {g \over 4} S_n^4\cr
&+ {g } R_n ( S_n^2 + S_{n-1}^2 + S_{n-1} S_n) \},\cr}}
with the defintions, $S_{-1} = 0$ and $R_0 = 0$. Morover we must take
$g > 0$, if the effective action (like the actual potential) is to be
bounded from below. This formalism provides a natural way to
investigate our set of multiple solutions, by removing the boundary
conditions on $S_0$ and $R_1$ and replace them with the asymptotic
condition that $R_n$ and $S_n$ are smooth functions as $n \rightarrow
\infty$. This is simply the one band ansatz in the extreme limit of
$n/N$ very large. Therefore it is again interesting to ask what is the
full set of local minima.

Earlier work on symmmetric solutions for the degenerate three-well
potential have observed the recursion coefficients with very
complicated, ``chaotic looking" behavior, when calculated by a
numerically method logically equivalent to minimizing an effective
potential\ref\CHAOS{M. Sasaki and H. Suzuki, Phys. Rev.  {\bf D43}
(1991) 4015; O. Lechtenfeld, Int. J. Mod. Phys. {\bf A7} (1992) 2335;
D. Senechal, Int. J. Mod. Phys. {\bf A7} (1992) 1491.}. We also have
observed complicated behavior for two degenerate wells when we allow
symmetry breaking terms ($S_n \ne 0$), which we have been able to
relate to the existence of our multiple solutions in the planar limit.
As an illustration consider the solution presented in
\fig\kinksoln{Graphs of recursion coefficients for a sponteneously
broken solution of the double-well potential. (a) The $R_n$ and (b)
the $S_n$ coefficients after 100,000 minimizaton steps from a random start.
(c) Orbit in the $A-B$ vs $C-D$ plane.  }.  However due to the
degeneracy of multiple solutions at $N_\infty$, great care must be
taken with the minization procedures.

For example we have minimized $V_{eff}$ for the double-well potential
with $N= 512$, $\mu = -2$, $g =1$ and $\sigma = 0.1$, starting from a
random distribution of 2048 coefficients for $R_n$ and $S_n$. Using a
variety of minimizaton procedures on the CM-5 at Boston University and
the NeXT station at CERN, we see that after only several 100
iterations the curves conform roughly with the large $N$ constraints but
they can have a great variety of coutours in the $A-B$ vs $C-D$ plane.
However if we go further for another 100,000 iterations, a smooth
spiral curve (see fig. 4 (c)) begins forming near critical $x$ ($x_{cr} =
1$), with the large $N$ constraints improving to about $1\%$ as might be
expected in a transition region from one of our $N=\infty$
solution to a particular double scaling solution with non-zero orbital
quantum number $l \ne 0$.  After 100,000 iterations the value of
$V_{eff}/N$ departs from its theoretical $N=\infty$ value  by
$0.0034$. The final results on questions as to the stability of non-zero
orbital solutions, the possibility of residual degeneracies at finite
$N$ and especially the existence of choatic regimes require accurate and
non-trivial compuatational power.  Further details on this as well as
a study of higher $1/N$ corrections will be presented in a future
publication\ref\BDJMT{ R.  Brower, N.  Deo, S.  Jain, P.  Mavromatis
and C-I Tan, in preparation.}.

\vfill\break\eject

\newsec{The Double Scaling Limit}

\subsec{Double Well $\phi^4$ Potential}

The double scaling equations  for the $\phi^4$ model
have been discussed by a number of authors \refs{\DSS, \DDJT, \CM {-}\HMPN}.
The steps involved in the double scaling analysis of symmetric breaking
solutions for a {\it $Z_2$ symmetric} potential are the same as that for
solutions of a general {\it asymmetric} potential \refs{\Nappi{-}\HMPN}, since
in both cases one includes both $R_n$ and $S_n$ in the analysis. One sets
 $x=1-\epsilon^2 t$, (recall $x=n/N$) and $\epsilon=N^{-1/3}$. For the
symmetric solution, $C_n=D_n=0$, while $A_n$  and $B_n$ are
\eqna\abdou$$\eqalignno{A_n&=a_0+\epsilon(f_e(t)+f_o(t))
+\epsilon^2(r_e(t)+r_o(t))+......&\abdou a\cr
B_n&=a_o+\epsilon(f_e(t)-f_o(t))+\epsilon^2(r_e(t)-r_o(t))+.....&\abdou b\cr}$$
On substituting this symmetric double scaling ansatz into the recursion
relations \rseqmug\ and equating terms with powers
$\epsilon^0,\epsilon^1,\epsilon^2,  and \epsilon^3$, we get eight equations,
two of
these are used up by $a_o$ ($\epsilon^0$ equations), the tree level result.
(Note $a_o=-\mu/(2g)$; in what follows, we adopt the convention where $\mu=-2$
and $g=1$). That leaves us with six unknowns and six equations hence all the
unknowns can be determined. Most of them are zero (e.g. $f_e=r_o=0....$), while
the others are determined in terms of $f_o(t)=f(t)$, {\it e.g.},
$r_e=(f^2-t)/4$. The function $f(t)$ satisfies the Painleve-II equation
\eqn\painii{f^{\prime\prime}-{1\over 4}~f^3+{1\over 2}~ft=0.} The suseptibility
$\chi\sim{\partial^2 \Gamma\over\partial \mu^2}\sim f^2/2-r_e= (f^2+t)/4$.

For the maximally asymmetric solution, the double scaling ansatz for $C_n$ and
 $D_n$
are
\eqna\cdou
$$\eqalignno{C_n&=\epsilon g(t)+\epsilon^2.....&\cdou a\cr
D_n&=-\epsilon g(t)+\epsilon^2.....&\cdou b\cr}$$
and  $A_n=a_0
+\epsilon^2r_e(t)+......$,
$B_n=a_o+\epsilon^2r_e(t)+\cdots$. Substituting this maximally asymmetric
double
 scaling
ansatz into the recursion relations and equating powers of $\epsilon $ we get,
$r_e=-(g^2+t)/4$, etc., and  \eqn\painasym{
 g^{\prime\prime}-{1\over 4}~g^3+{1\over 2}~gt=0,}
the same as Eq. \painii, with $g$ replacing $f$. Under this ansatz, the
suseptibility $\chi$ can be expressed as  $(g^2+t)/4$.

We next consider the general symmetry breaking solutions, where $A-B\neq
0$ and $C-D\neq 0$ in the planar limit. Substituting  the double
scaling ansatz, Eqs. \abdou{}\ and \cdou{}\ into the recursion relations and
equating powers of $\epsilon $ we get $r_e=(f^2-g^2-t)/4$, etc., and the
following coupled equations
\refs{\Nappi{-}\HMPN}
\eqna\coupledeq
$$\eqalignno{f^{\prime\prime}-f(g^2+f^2)/4+ft/2&=0,&\coupledeq a\cr
g^{\prime\prime}-g(g^2+f^2)/4+gt/2&=0.&\coupledeq b\cr}$$
 The suseptibility is now
given by  $\chi=( f^2+g^2+t)/4$.

While Eqs. \coupledeq{}\ have been obtained previously in the context of
asymmetric potentials, we would like to emphasize that they describe the
multiple (and in general symmetry breaking) solutions that exist even for a
$Z_2$-symmetric potential. To see the symmetry breaking nature of these
solutions more explicitly and to make contact with the tree level discussion in
the previous section, introduce a two-dimensional vector ${\vec r}=(g,f)$, in
terms of which the coupled equations can be written as $\ddot{\vec r}-(1/4)(
r^2-2t){\vec r}=0$. We can next make a change of coordinates %
\eqn\fgrt{f=r~sin\theta(t), \hskip40pt
                   g=r~cos\theta(t),}
so that  $\chi\sim (r^2+t)/4$ and the coupled equations become %
\eqna\painrl$$\eqalignno{{\ddot
r}-{1\over4}~r^3+{1\over2}~rt-{l^2\over {r^3}}&=0,&\painrl a\cr
 r^2 {\dot \theta}&=l. &\painrl
b\cr}$$
Note that since $A-B\propto f$ and $C-D\propto g$ in the double scaling limit,
the variable $\theta$ is the same as that introduced in the previous section.
The constant $l$ in Eq. \painrl{b}\ is the ``angular~ momentum'', and it is a
constant of the motion due to the $U(1)$ invariance of Eqs. \coupledeq{}. Note
that  for $l=0$ equation \painrl{a}\ is just the Painleve-II equation in the
$r$
coordinate. Thus in the $l=0$ sector of this model, we reproduce the same
double
scaling results for both the symmetric and maximally asymmetric solutions.  But
for the $l\not=0$ sector the double scaling equation is  different; hence the
behavior of the system in this sector for the multiple solutions is different.
Iterating Eq. \painrl{a}\   at large $t$, one finds
\eqn\free{\chi= {{3t\over 4}
-({1+l^2\over 4})t^{-2}+\cdots}}

At the tree level we had a degeneracy of solutions parametrized by the
function $\theta(x)$, whose value could be independently chosen for
$x\epsilon [0,1]$. The double scaling analysis based on the  \abdou\ and
\cdou\ tells us that only a two parameter family of these solutions, labelled
by $l$ and one global rotation angle $\theta_0$, survives in the
double scaling limit. ($\theta(x)$ is no longer any function of $x$,
but constrained such that $r^2 {\dot \theta}=l$ is a constant.)
Solutions labelled by different values of $l$ give rise to the same
susceptibility at tree level (the first term in \free\ is
$l$-independent) as expected, but differ at higher orders.

A relevent question that arises is: Just as the symmetric solution ($S_n=0$) is
the ``natural" solution that follows from the symmetric potential $V(x)$,
(``natural" in that it respects the $Z_2$ symmetry of $V(x)$, and the recursion
coefficients are specified by initial conditions given by the integrals
 discussed
in section 2), is there a perturbed potential of which the symmetry breaking
solutions are natural solutions? This is presently being investigated. At this
point we remark that the perturbation cannot be a rigid translation of the
potential, which induces linear and cubic terms. Such a perturbation has been
discussed in Ref. \HMPN. It can be easily seen that the natural solution to the
shifted potential $V_b(x)\equiv V(x-b)$ is given by the same $R_n$ as for
$V(x)$, with $S_n=b={\rm const.}$

 Should one decide to
introduce a small explicit symmetry-breaking term, $\sigma =N^{-{2/3}}\tilde
\sigma$,  as was done in Ref. \Nappi, the vector equation above remains $U(1)$
invariant by simply adding a ``constant magnetic field" term of the form
$B_0{\dot{\vec r}}\times\hat z$, where $B_0\propto \tilde\sigma$ and $\hat{z}$
denotes a unit normal to this two-dimensional plane. One can thus again reduce
it to a single radial equation with the solution depending on a ``generalized
angular mumentum", $l=r^2 {\dot \theta}+B_0r^2/2={\rm constant}$,
\eqn\painrltwo{{\ddot r}-{1\over4}~r^3+{1\over2}~rt-{(l-{B_0r^2\over
2})^2\over{r^3}}=0.}

\subsec{The Gaussian Penner Model}

The double scaling solutions for the Gaussian Penner model have been
discussed in Refs. \Tii\ and \CDL. We reproduce the proofs below for
completeness (also simplifying them somewhat). The critical point is $t=-1$. By
strictly enforcing $Z_2$ symmetry, this model can be solved exactly first at
$t>0$, so that the criticality at $t=-1$ can be exhibited explicitly.  Note
that   $\langle n|{\hat \phi}^{-1}|n\rangle=0$ by $Z_2$ symmetry, Eq.
\rseqpen{b}\ thus reduces to  $S_n=0$, the symmetric ansatz. Eq. \rseqpen{a}\
can also be solved exactly for all n. Since $W_n=\mu R_n $ and  since $\langle
n-1|{\hat \phi}^{-1}|n\rangle=0$ for  n even, $\langle n-1|{\hat
\phi}^{-1}|n\rangle={1\over \sqrt{R_n}}$ for n odd, it follows that, for n
even,
$R_n={n/ {\mu N}}$, and for n odd, $R_n=(n+tN) /{\mu N}$\Tii.

 Since we know the
exact result for the $R_n 's$ for the symmetric solution, the exact free energy
may be obtained  %
\eqn\freepensym{\Gamma=\sum_{k=1}^{N/2-1}
 ~k~log\left[(2k+\mu+1)(2k+\mu-1)\right],}
where $t=-1+{\mu\over N}$. On expanding the free energy in powers
of $\mu$ we get
\eqn\fsymmu{\Gamma={1\over
4}~\mu^2~log~\mu+{1\over12}~log~\mu+....} The coefficient of the second
$\log\mu$, $\chi_1=1/{12}$, comes from the torus contribution, indicates that
this free energy  {\it cannot} be identified with
 the Legendre transform of the free energy of the c=1 string at self dual or
twice the self dual radius \Tii.

It has been stressed in Ref. \Tii\ that  the exact solution to the
Gaussian Penner model  is characterized by the fact that $B(x)$ has a {\it
linear}
zero at $x=1$ when $t=-1$ while $A(x)$ is non-zero there, (see Eq. \absympen).
This same feature also   holds in general for symmetry breaking
solutions where $C(x)=-D(x)\neq 0$. However, the maximally asymmetric solution
provides an exception to this rule. When $A(x)=B(x)$, one has
$A(x)=B(x)\sim (x+t)^{1/2}$, (see Eq. \acasym). That is, both $A(x)$ and
$B(x)$ have  {\t square-root} type behavior near $x=1$ in the spherical limit
at
$t=-1$. Since it is the behavior of $R_n$ near $x=1$ which determines the
criticality of the model, it follows that the resulting double scaling limit
for the maximally asymmetric ansatz could be { non-generic}.

Let us next concentrate on the maximally asymmetric solution. We note first
that, with $S_n\neq 0$,   Eqs. \rseqpen{a}\ and \rseqpen{b} can be re-written
as \Ti\Tii\
\eqna\rseqpentwo$$\eqalignno{W_n+W_{n+1}+S_nY_n&={2n+1+Nt\over
N},&\rseqpentwo a\cr S_N[W_{n+1}-W_n-{1\over
N}]&=R_nY_{n-1}-R_{n+1}Y_{n+1},&\rseqpentwo b\cr}$$
so that matrix elements of $\hat \phi$ would not appear explicitly.
For the maximally asymmetric solution the double scaling
solutions may be found as follows. With $x=1-z/N$, $t=-1+\mu/N$,
$\epsilon=1/\sqrt{N}$,  $A_n, C_n,~ {\rm and}~ D_n$
can be expanded as \CDL\
 \eqn\acdpenasy{\eqalign{A_n&=1/\sqrt{N}~\rho(z)+...\cr
C_n&=1+1/\sqrt{N}~\sigma(z)+1/N~\sigma_1(z)+....\cr
D_n&=-\left(1+1/\sqrt{N}~\sigma(z)+1/N~\sigma_1(z)+....\right).\cr}}
On
substituting into the recursion relations \rseqpentwo\ and equating equal
powers of $N$, we can determine $\sigma(z)$ in terms of $\rho(z)$ and
$\sigma_1(z)$ in terms of $\sigma(z)$. The equation for $\rho$ is
$\rho(z)\rho(z-1)=\mu-1/2+z$ and the double scaling result for $R_n$
is  $R_n\sim \Gamma \left(\half(N-n+\mu+3/2) \right)/
 \Gamma\left(\half(N-n+\mu+1/2) \right) $. The double scaled free energy
 is $\Gamma=\sum_{k=1}^{N/2-1} ~k~log
\left[(2k+\mu+1/2) (2k+\mu-1/2)\right]$ plus$ \mu$  independent  terms. On
expanding in powers of $\mu$ one gets
\eqn\fasympenmu{\Gamma={1\over4}~\mu^2~log~\mu-{5\over 48}~log~\mu.....}
The coefficient $\chi_1=-5/{48}$ confirms that this criticality corresponds to
that for the free energy of the c=1 string at twice the self
dual radius as conjectured in \DV.  We  note  that although the tree level free
energy gave identical results for the symmetric and maximally asymmetric ansatz
(see remarks below Eq. \acasym), the double scaled free energies \fsymmu\ and
\fasympenmu\  are quite different.

\vfill\eject
\newsec{Discussion}

The existence of multiple double scaling solutions at
the same critical point, which share the same tree level behavior, is an
unusual and previously unnoticed feature. We have exhibited this
behaviour in two completely different models-the double-well $\phi^4$
potential and the Gaussian Penner model. Both these models possess
$Z_2$ symmetry, and a charateristic feature of the class of solutions
we consider is the fact that only one of the solutions respects this
symmetry to all orders, all other solutions break this symmetry.
Although the physical consequences of symmetry breaking in matrix
models are not yet fully understood, it is useful to consider the
analogy of the multiplicity of solutions here to the property of
coexisting ferromagnetic phases below the Curie point.  In
the Ising model there is an infinite set of mixed phases with
identical free energy per unit volume in the infinite volume limit.
The domain walls that characterize the mixed phases give rise to lower
order contributions in the free energy expanded around the infinite
volume limit. The ``bulk'' contribution in the multiple solutions is the same
(the tree level eigenvalue density is the same) but they differ by amounts
suppressed by powers of $1/N$, like ``surface'' contributions.

A significant difference between the Gaussian Penner model and the
symmetric $\phi^4$ model is that in the Gaussian Penner model there
does not seem to be an angular momentum parameter $l$ characterizing
the double scaling solutions.  Further, the double scaled free
energies of the symmetric and maximally asymmetric solutions already
differ, unlike in the $\phi^4$ case where both of these were $l=0$
solutions with the same double scaled free energy. In spite of these
differences, however, both models display the same general phenomenon,
namely, that the enlarged class of symmetry breaking solutions
produces the {\it same} free energy at {\it tree level}, and contain
solutions that produce {\it different} free energies at {\it higher
orders}. We expect this to be a generic feature of multi-cut matrix
models.  The existence of multiple solutions is related to the fact
that when the potential has more than one minima, the number of smooth
functions required to represent the recursion coefficients exceeds the
number of constraints obtained from the recursion relations. Thus
multiple solutions will exist even when the potential has no symmetry
as we noted for $\sigma \ne 0$ in the $\phi^4$ model.

It would be interesting to know whether and how these solutions can
tunnel into each other.
This would be particularly interesting for the Gaussian Penner model where
the maximally asymmetric solution corresponds to the $c=1$ string compactified
at twice the self-dual radius.
We are now studying in greater detail higher
order terms beyond the planar approximations as well as numerical
solutions at finite N to the effective potential, Eq. \effect, to
determine more precisely the relationship between multiple solutions
of the planar versus the double scaling limit.

{\bf Acknowledgements }

We would like to thank Robert Edwards and Panayotis Mavromatis for
many helpful discussions through out this research and greatfully
acknowledge the supportive atmosphere in the CERN theory group during
the visits of RCB and CIT and the Center for Computational Science at
Boston University.  The work was supported in part by the DOE grants
DE-AC02-89ER40509 and DE-AC CO76ER03130.A021-Task A, the NSF grant
PHY-87-14654 and the Packard Foundation.  Part of this work was done
while one of us (CIT) was on sabbatical leave at LPTHE-Orsay, France.

\listrefs

\listfigs

\end